\documentclass[aps,prl,amsfonts,amsmath
,twocolumn,showpacs
,nofootinbib
]
{revtex4-1}

\usepackage{epsfig}
\usepackage{graphicx}
\usepackage{mathptmx}      % use Times fonts if available on your TeX system
\usepackage{latexsym}
\usepackage{bm}

\def\slashchar#1{\setbox0=\hbox{$#1$}
   \dimen0=\wd0 \setbox1=\hbox{/} \dimen1=\wd1
   \ifdim\dimen0>\dimen1 \rlap{\hbox to \dimen0{\hfil/\hfil}} #1
   \else  \rlap{\hbox to \dimen1{\hfil$#1$\hfil}} / \fi}

\def\tr{{\rm tr}}
\def\Tr{{\rm Tr}}
\newcommand{\MeV}{\,{\rm MeV}}

\newcommand{\SU}{{\rm SU}}
\newcommand{\U}{{\rm U}}
\newcommand{\Eq}[1]{Eq.~(\ref{eq:#1})}

\newcommand{\vx}{{\bm{x}}}

\newcommand{\vA}{{\bm{A}}}
\newcommand{\ignore}[1]{}

\begin{document}

\title{The Polyakov loop and the hadron resonance gas model}

\author{E. Meg\'{\i}as}
\email{emegias@ifae.es}

\affiliation{Grup de F\'{\i}sica Te\`orica and IFAE, Departament de F\'{\i}sica, Universitat Aut\`onoma de Barcelona, Bellaterra E-08193 Barcelona, Spain}

\author{E. \surname{Ruiz Arriola}}
\email{earriola@ugr.es}

\author{L. L. Salcedo}
\email{salcedo@ugr.es}

\affiliation{Departamento de F{\'\i}sica At\'omica, Molecular y Nuclear \\
and
  Instituto Carlos I de F{\'\i}sica Te\'orica y Computacional, \\ Universidad
  de Granada, E-18071 Granada, Spain.}

\date{\today}

\begin{abstract}
The Polyakov loop has been used repeatedly as an order parameter in the
deconfinement phase transition in QCD. We argue that, in the confined phase,
its expectation value can be represented in terms of hadronic states,
similarly to the hadron resonance gas model for the pressure. Specifically,
$L(T) \approx \frac{1}{2}\sum_\alpha g_\alpha \,e^{-\Delta_\alpha/T}, $ where
$g_\alpha$ are the degeneracies and $\Delta_\alpha$ are the masses of hadrons
with exactly one heavy quark (the mass of the heavy quark itself being
subtracted). We show that this approximate sum rule gives a fair description
of available lattice data with $N_f=2+1$ for temperatures in the range
$150\MeV<T<190\MeV$ with conventional meson and baryon states from two
different models. For temperatures below $150\MeV$ different lattice results
disagree. One set of data can be described if exotic hadrons are present in
the QCD spectrum while other sets do not require such states.
\medskip
~\\ 
\end{abstract}


%\keywords{finite temperature; heavy quarks; chiral quark models; Polyakov Loop}

\maketitle

%\tableofcontents

%\section{Introduction}
%\label{sec:introduction}

{\em Introduction.---}The transition from the hadronic phase to the
quark-gluon plasma phase has been a recurrent topic in hadronic
physics~\cite{Fukushima:2011jc}. In pure gluodynamics or, equivalently, for
infinitely heavy quarks, this is a true phase transition. The order parameter
is identified as the thermal Wilson line or Polyakov loop
\cite{Polyakov:1978vu,Susskind:1979up,Svetitsky:1985ye},
\begin{equation}
\Omega = {\sf P} e^{i \int_0^{1/T}\! A_0\, dx_0}
,\quad
L(T) =  \langle  \tr \,\Omega \rangle
,
\end{equation}
where $A_0 $ is the gluon field, $T$ is the temperature and ${\sf P}$ denotes
path ordering. $L(T)$ changes abruptly from zero to near $N_c$ (the number of
colors) due to the breaking of the center symmetry $\mathbb{Z}(N_c)$ for $T
> T_c$.  In QCD, i.e., for dynamical quarks, the center symmetry is
explicitly broken by the quarks and one has instead a smooth crossover
\cite{Aoki:2006we} and the critical temperature $T_c$ is usually defined by
the condition $L^{\prime\prime}(T_c)=0$. Lattice simulations show that chiral
symmetry is restored when quarks and gluons are deconfined. These theoretical
insights strongly suggested the experimental quest for the quark-gluon plasma
in current facilities. The Polyakov loop also serves as a gluonic effective
degree of freedom in the successful Polyakov--Nambu--Jona-Lasinio (PNJL) and
Polyakov--quark-meson models (PQM) to describe hot and/or dense QCD
\cite{Meisinger:1995ih,Fukushima:2003fw,Megias:2004hj,Ratti:2005jh,Schaefer:2007pw}.

Since hadrons (and possibly glueballs) are the physical states in the confined
phase it should be expected, by quark-hadron duality, that physical quantities
admit a representation in terms of hadronic states. The QCD pressure presents
a prime example of this, through the hadronic resonance gas model (HRGM)
\cite{Hagedorn:1984hz, Agasian:2001bj,Tawfik:2004sw, Megias:2009mp,
  Huovinen:2009yb,Borsanyi:2010cj, Bazavov:2011nk},
\begin{equation}
\frac{1}{V}\log Z = 
-\int \frac{d^3 p}{(2\pi)^3} \sum_\alpha \zeta_\alpha g_\alpha 
\log \left( 1 - \zeta_\alpha e^{-\sqrt{p^2+M_\alpha^2}/T} \right) \,,
\label{eq:rgm}
\end{equation}
with $g_\alpha$ the degeneracy factor, $\zeta_\alpha=\pm 1$ for bosons and
fermions respectively, and $M_\alpha$ the hadron mass. These resonances are
 the low-lying states listed in the review by the Particle Data Group
(PDG)~\cite{Nakamura:2010zzi}. Actually, in the large $N_c$ limit this
expectation becomes a true theorem in QCD since the flavor resonances become
narrow $\Gamma/M=O(1/N_c)$ (see
e.g. \cite{Cohen:2004cd,Masjuan:2012gc}). After some controversies, lattice
calculations seem to suggest that this is also a good approximation in the
physical case $N_c=3$ \cite{Borsanyi:2010bp}. This problem has been addressed
within a strong coupling expansion for heavy quarks in
\cite{Langelage:2010yn}.

Despite its prominent theoretical role, $\Omega(\vx)$ does not appear to be
directly accessible in the laboratory, being most naturally defined in the
imaginary-time formalism of field theory at finite temperature
\cite{Matsubara:1955ws,Landsman:1986uw} (see \cite{MoralGamez:2011en} for its
definition within the real-time formalism). The realization of the Polyakov
loop through a static (or heavy) quark and its relation with the heavy-quark
self-energy (or even with the binding energy between a static and a dynamical
quark \cite{Petreczky:2012rq}), is not new
\cite{Svetitsky:1985ye,Kaczmarek:2002mc}, but the phenomenological
consequences of this fact have not yet been extracted at a quantitative level.
Here we argue that the Polyakov loop in the confined phase can also be
represented in terms of hadronic properties in a direct and quantitative way,
similarly to the HRGM for the pressure.

{\em Polyakov loop and hadronic spectrum.---} In the Hamiltonian formulation
of QCD~\cite{Kogut:1974ag,Luscher:1976ms}, the gauge is partially fixed by the
condition $A_0=0$ and the dynamical degrees of freedom are contained in the
spatial gluons $\vA$ and the quarks.  The time-independent gauge
transformations $g(\vx)$ are still a residual symmetry of the Hamiltonian
acting in the Hilbert space $\mathcal H$ of functionals
$\Psi(\vA,q,\bar{q})$. The gauge group $\SU(N_c)$ decomposes $\mathcal H$ into
invariant subspaces labeled by an irrep $r$ at each point $\vx$: $ \mathcal{H}
= \bigoplus_{\{r(\vx)\}} \mathcal{H}_{\{r(\vx)\}} $.  In the Euclidean lattice
formulation, the role of the integration over $A_0$, or equivalently
integration over $\Omega(\vx)$ with the Haar measure, is to project onto the
physical subspace, which requires a color singlet at every point $\vx$.

An infinitely heavy quark (of a new flavor) sitting at $\vx_0$ is a spectator
with spin and color degrees of freedom only. For the gluons and dynamical (as
opposed to heavy) quarks, this is equivalent to living in the subspace
$\mathcal{H}_{r(\vx_0)={\bf 3}}$: a color singlet at every point except
$\vx_0$, which is in the fundamental representation (${\bf 3}$ for three
colors). The projector onto this subspace is obtained by adding the factor
$N_c\tr(\Omega(\vx_0))$ to the Haar measure \cite{Tung:1985bk}.  For the
expectation value of the Polyakov loop this immediately implies the relation
\cite{Luscher:2002qv,Jahn:2004qr}
\begin{equation}
L_{\rm bare}(T)
= 
\frac{1}{2}\frac{\Tr_{h,\vx_0} (e^{-H/T})}{\Tr_{\rm phys} ( e^{-H/T})}
\,.
\label{eq:31}
\end{equation}
The factor $N_c$ in the projector represented the trivial degeneracy of the
system formed by gluons plus dynamical quarks in the fundamental
representation at $\vx_0$, and is canceled when this is combined with the
spectator quark to form a color singlet. The factor $1/2$ removes the double
counting from the two spin states of the spectator quark.  The l.h.s. is
independent of the heavy quark spin (as the Polyakov loop carries no spin) and
this is fully consistent with the well-known heavy-quark spin symmetry present
in QCD \cite{Isgur:1989vq,Neubert:1993mb}.  The (infinite) mass of the
spectator quark is not included in $H$.

\Eq{31} is exact for the bare Polyakov loop and the partition
functions in $\mathcal{H}_{{\rm phys}}$ and $\mathcal{H}_{r(\vx_0)=3}$ on the
lattice. In the renormalized continuum limit the relation still holds, after
removing the additional specific UV divergence introduced by the heavy quark
self-energy in $L(T)$ and $\Tr_{h,\vx_0} (e^{-H/T})$. Such removal leaves a
nonperturbative ambiguity by an additive constant in the Polyakov loop free
energy $F(T)=-T\log L(T)$
\cite{Kaczmarek:2002mc,Dumitru:2003hp,Gupta:2007ax,Gavai:2010qd,%
Borsanyi:2010bp,Mykkanen:2012ri}.

The (renormalized) partition functions in \Eq{31} are saturated by states of
the spectrum. Since the spectator quark can be reached smoothly by taking the
infinite mass limit of a heavy quark at rest, this implies
\begin{equation}
L(T)
=
\lim_{m_h\to\infty}
\frac{1}{2}\frac{\sum^\prime_\alpha g_{h\alpha}  e^{-(E_{h\alpha}-m_h)/T}}
{\sum_\alpha g_\alpha  e^{-E_\alpha /T}} \,,
\label{eq:32}
\end{equation}
where $m_h$ denotes the heavy quark mass. Here, the sum in the denominator is
just the QCD partition function and so it includes all possible states made of
gluons and dynamical quarks (labeled by $\alpha$). On the other hand, the sum
in the numerator includes all possible QCD states with exactly one heavy
quark $h$ at rest,\footnote{Such statement would be meaningless for dynamical
  quarks, but not for infinitely heavy quarks.} plus gluons and dynamical
quarks (jointly labeled by $h\alpha$). The difference $E_{h\alpha}-m_h$
explicitly removes the heavy quark mass from the total energy of the state.

For temperatures well below the crossover, we expect the previous states to be
of hadronic type (and possibly glueballs). In particular, the heavy quark $h$
will form a hadron with the dynamical quarks, typically, a meson of hybrid
type, i.e., formed by the heavy quark and a dynamical antiquark, $h\bar{q}$,
or a hybrid baryon with the heavy quark and two dynamical quarks, $hqq$.

The HRGM for the QCD partition function follows naturally from assuming that
the QCD interaction primarily confines quarks into hadrons and that purely
hadronic interactions can be neglected. Under this assumption, the numerator
of \Eq{32} contains one hybrid heavy-light hadron at rest plus exactly the
same multi-hadron states occurring in the denominator. This yields a
cancellation between numerator and denominator. Therefore, within the same
approximations leading to the HRGM, we expect the following relation to hold
between the Polyakov loop expectation value in the confined phase and the
hadronic spectrum
\begin{equation}
L(T) \approx
\frac{1}{2}\sum_\alpha \!{}^{{}^{\prime\prime}} g_{h\alpha}\,  e^{-\Delta_{h\alpha}/T} \,,
\quad
\Delta_{h,\alpha} = M_{h\alpha}-m_h \,.
\label{eq:34}
\end{equation}
Here the sum is over all states made just of a single hybrid hadron at rest
(with exactly one heavy flavor quark, mass $M_{h\alpha}$ and degeneracy
$g_{h\alpha}$), and no additional hadrons.

Of course, neither the sum rule in \Eq{34} nor the HRGM can be accurate when
unconfined states of the spectrum start to play a role, that is, for
temperatures in the crossover region or above.  For instance, $L(T)$ is a
decreasing function at high enough temperatures, in the perturbative regime
\cite{Gava:1981qd}, while the Boltzmann distribution in the r.h.s of \Eq{34}
is always increasing as a function of $T$. On the other hand, the mass of the
heavy-light hadron is an observable (a renormalization invariant) but depends
on the heavy flavor, while $\Delta_{h\alpha}$ is universal in the heavy quark
limit but has some running from $m_h$, which is itself not an observable. The
Polyakov loop is renormalization invariant and well defined, modulo the
abovementioned shift ambiguity in the free energy. This can be compensated by
a corresponding shift in the heavy quark mass.

\begin{figure}[t]
\begin{center}
\epsfig{figure=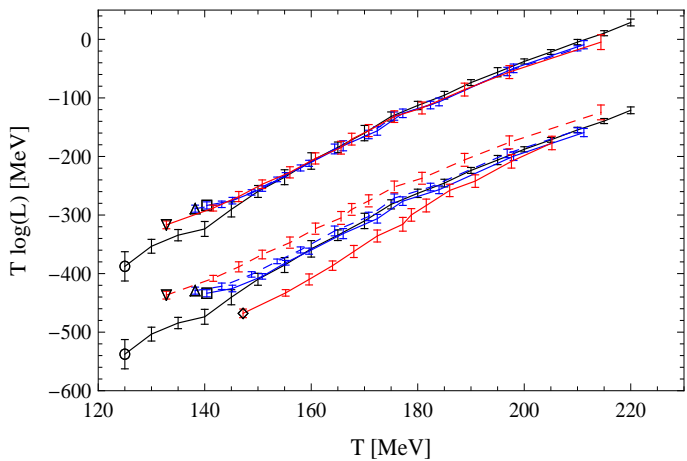,height=6cm,width=9.0cm}
\end{center}
\caption{(color online). $T \log(L(T))$ as a function of $T$ (units in MeV)
  from simulations on the lattice with $2+1$ physical dynamical quark
  masses. Lower set of lines: data after a common shift $C=-150\MeV$ (just
  for displaying purposes) for continuum extrapolated stout (black solid line,
  ``circle'') \cite{Borsanyi:2010bp}, HISQ/tree action $N_t=12$ scale set
  $r_1$ (blue solid line, ``square'') and $f_k$ (blue dashed line, ``up
  triangle''), and asqtad action $N_t=12$ scale set $r_1$ (red solid line,
  ``down triangle'') and $f_k$ (red dashed line, ``rhombus'')
  \cite{Bazavov:2011nk}. Upper set of lines: same data (except asqtad
  scale set $r_1$) with shifts $C=0$, $0$, $-10\MeV$ and $-30\MeV$, for stout,
  HISQ/tree scale $r_1$, HISQ/tree scale $f_k$, and asqtad scale $f_k$,
  respectively.  }
\label{fig:1}
\end{figure}

\begin{figure}[t]
\begin{center}
\epsfig{figure=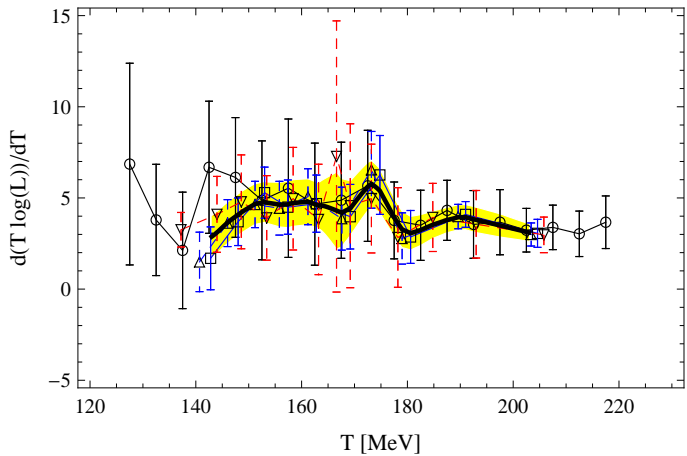,height=6cm,width=9.0cm}
\end{center}
\caption{(color online). Extraction of the slope $\frac{d}{dT}(T\log(L(T)))$
  (dimensionless) as a function of $T$ (in MeV) from the same four lattice
  data sets of Fig.~\ref{fig:1}. The error bars in the slope were obtained
  from assuming a linear interpolation between measured points. The black
  thick line indicates the average in the range of temperatures common to the
  four sets. The yellow strip indicates the uncertainty.}
\label{fig:2}
\end{figure}

{\em Estimates from the physical spectrum.---}In order to compare different
Polyakov loop determinations with the hadronic sum rule, they have to be
brought to a common renormalization condition. Two such determinations are
related by $L^\prime(T)= e^{\,C/T}L(T)$, for some constant energy shift
$C$. In Fig.~\ref{fig:1} we compile five Polyakov loop data sets, obtained
with physical quark masses and three flavors on the lattice
\cite{Borsanyi:2010bp,Bazavov:2011nk}. The plot shows that four of them agree
after applying suitable finite renormalizations (no attempt has been made to
optimize the agreement), in a wide range of temperatures. Unfortunately, the
agreement deteriorates at lower temperatures, below $T\approx 150\MeV$. This
region is relevant for comparison with the hadronic sum rule. Since there is
no ``true'' value of $L(T)$, a finite renormalization, or choice of heavy
quark mass, should be admitted too in the hadronic sum rule. Nevertheless,
large renormalizations (compared with the hadronic scale) would be unnatural,
and would probably signal an inadequacy of the sum rule, or of the
renormalization prescription used for the Polyakov loop. A neat way to remove
any ambiguities is to work with the derivative of $T \log(L(T))$ with respect
to $T$. This slope is sensitive to the effective number of states at a given
temperature. Although with some noise, Fig.~\ref{fig:2} shows that a signal
can be extracted in this way.

A natural step is to check to what extend the hadronic sum rule is fulfilled
by experimental states compiled in the PDG. Several sources of error should be
kept in mind when doing this, among other, that not all needed states may have
been compiled, that the heavy quarks in nature have a finite mass, that their
current mass is scale dependent, and that the quark masses on the lattice may
not be identical to the physical ones. Hadrons with a bottom quark would be
optimal, due to the large quark mass compared to $\Lambda_{\rm QCD}$, but the
available data are scarce, so we turn to charmed hadrons. Specifically, we
consider the lowest lying single-charmed mesons and baryons with $u$, $d$, and
$s$ as the dynamical flavors, with quarks in relative $s$-wave inside the
hadron. For mesons, these are usually identified with the states (spin-isospin
multiplets) $\bar{D}$, $\bar{D}_s$, $\bar{D}^*(2010)$ and $\bar{D}_s^*$, and
for baryons, with $\Lambda_c$, $\Sigma_c(2455)$, $\Xi_c$, $\Xi_c^\prime$,
$\Omega_c$, $\Sigma_c(2520)$, $\Xi_c(2645)$, and $\Omega(2770)$. A total of 12
meson states and 42 baryon states.  \ignore{ $IJ^\pi=\frac{1}{2}0^-$
  $\bar{D}$, the $00^-$ $\bar{D}_s$, the $\frac{1}{2}1^-$ $\bar{D}^*(2007)$
  and the $01^-$ $\bar{D}_s^{*}$. For baryons, with the states
  $0\frac{1}{2}^+$ $\Lambda_c$, the $1\frac{1}{2}^+$ $\Sigma_c(2455)$, the
  $\frac{1}{2}\frac{1}{2}^+$ $\Xi_c$, the $\frac{1}{2}\frac{1}{2}^+$
  $\Xi_c^\prime$, the $0\frac{1}{2}^+$ $\Omega_c$, the $1\frac{3}{2}^+$
  $\Sigma_c(2520)$, the $\frac{1}{2}\frac{3}{2}^+$ $\Xi_c(2645)$, and the
  $0\frac{3}{2}^+$ $\Omega(2770)$. For the charm quark mass, to be subtracted
  from the hadron masses, we take the $\overline{\rm MS}$ scheme value
  $m_c=1290^{+50}_{-110}\MeV$, also directly from~\cite{Nakamura:2010zzi}.}

\begin{figure}[t]
\begin{center}
\epsfig{figure=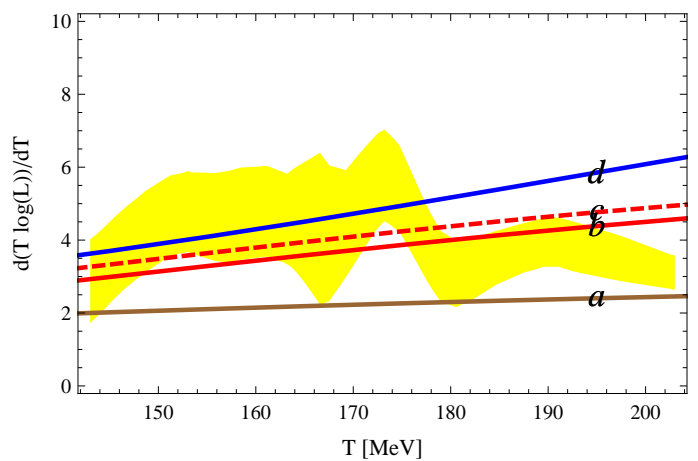,height=6cm,width=9.0cm}
\end{center}
\caption{(color online). Comparison of $\frac{d}{dT}(T\log(L(T)))$ (yellow
  strip) with $\frac{d}{dT}(T\log(\frac{1}{2}\sum^{\prime\prime}_\alpha
  g_{h\alpha} e^{-\Delta_{h\alpha}/T}))$ from hadronic states (mesons plus
  baryons): lowest-lying hadrons from PDG (solid brown line, label $a$), RQM
  states from \cite{Godfrey:1985xj,Capstick:1986bm} with quark $c$ (solid red
  line, label $b$), and with quark $b$ (dashed red line, label $c$), and bag
  model estimate including states up to $\Delta=5500\MeV$ (solid blue line,
  label $d$).}
\label{fig:3}
\end{figure}

The plot in Fig.~\ref{fig:3} shows that the lowest-lying states fall short to
saturate the sum rule, regardless of the choice of mass of the charmed quark,
$m_c$. This is not surprising as any model predicts many excited states on top
of the lowest-lying ones, as is also the case for light-quark hadrons. Adding
more states from the PDG does not seem practical due to the fragmentary
information available. Instead we turn to hadronic models. The aim is not so
much to have a detailed description of the various states but to give a
sufficiently good overall description of the whole spectrum. To this end we
consider the relativized quark model (RQM)
\cite{Godfrey:1985xj,Capstick:1986bm}, and the bag model
\cite{Chodos:1974je,Bernotas:2008bu}.  We have verified that the RQM provides
a good account of the trace anomaly in \cite{Borsanyi:2010bp}. The total
number of hadron states computed in \cite{Godfrey:1985xj,Capstick:1986bm} with
one $c$ quark is $117$ for mesons and $1470$ for baryons, corresponding to a
maximum value of $\Delta=M-m_c$ about $1500\MeV$. For hadrons with one $b$
quark, $87$ mesonic and $1740$ baryonic states, with a similar upper bound for
$\Delta$. In both cases we have supplemented missing states with strange
quarks by means of the equal spacing rule \cite{Savage:1995dw} and a $s$ quark
mass of $109\MeV$ (extracted from the lowest-lying hadrons masses). The
prediction based on these hadronic states is displayed in
Fig.~\ref{fig:3}. The two sum rules are closer to the Polyakov loop result but
still tend to stay below it in the range $T<175\MeV$, a consequence of the
truncation of states to $\Delta<1500\MeV$. It is noteworthy that the bottom
sum rule gives a better value, as it would be expected due to the larger mass
of the $b$ quark.

The other model we consider is more schematic but allows us to easily include
a larger number of states. This is a simplified MIT bag
model \cite{Chodos:1974je,Bernotas:2008bu}, in order to correctly count the
number of states without fine details such as multiplet splittings. As bag
energy we take
\begin{equation}
\Delta=\frac{\sum_i {n_i\omega_i}-Z}{R}+\frac{4\pi}{3}R^3 B+\sum_i m_i,
\end{equation}
where $n_i$, $\omega_i$ and $m_i$ are the occupation number, bag frequency and
current quark mass ($m_u=m_d=0$, $m_s=109\MeV$). The model gives directly
$\Delta$ after adjustment of $R$, without center of mass corrections, since
the heavy quark has actually infinite mass (not included), sits at the center
and plays no active role. The sum over $i$ runs over just one antiquark for
$h\bar{q}$ mesons and two quarks for $hqq$ baryons. We set $B=(166\MeV)^4$
\cite{Bernotas:2008bu}, and find $m_c=1390\MeV$ and $Z=0.5$ from a fit to the
single-charmed lowest-lying mesons and baryons masses. The result for the bag
model with $\Delta<5500\MeV$ is displayed in Fig.~\ref{fig:3}. The more
complete description of the hadronic spectrum gives a better account of the
Polyakov loop data in the range $145\MeV < T < 175\MeV$. We have checked that:
i) For this range of temperatures heavier states become irrelevant for the sum
rule. We see this by projecting the cumulative number of states assuming a
power law dependence for mesons and for baryon \cite{Kapusta:1981ue}. ii)
Truncation to $\Delta<1500\MeV$ gives a result quite consistent with that of
RQM. And iii) A similar power-law projection of the spectrum for the RQM, to
estimate the effect of adding the states above $\Delta=1500\MeV$, also
reproduces quantitatively the bag model result. Fig.~\ref{fig:3} also shows
that the hadronic sum rule eventually overshoots the Polyakov loop result, as
the crossover to the deconfined phase sets in. This mimics the same behavior
in the HRGM \cite{Bazavov:2009zn,Borsanyi:2010bp}.

\begin{figure}[t]
\begin{center}
\epsfig{figure=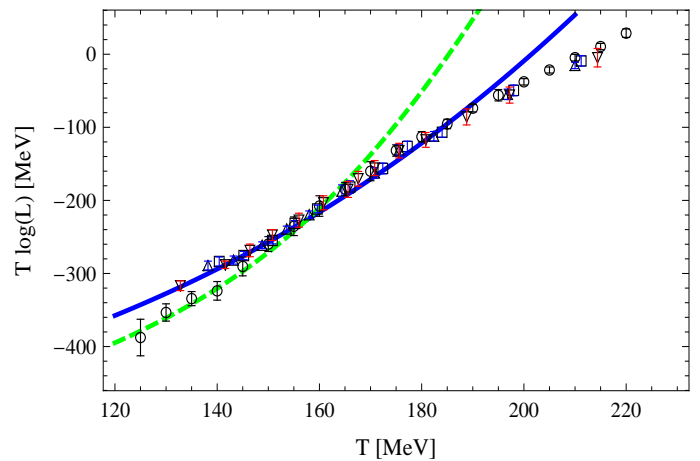,height=6cm,width=9.0cm}
\end{center}
\caption{(color online). $T\log(L(T))$ (in MeV) for the four lattice data sets
  of Fig.~\ref{fig:1} compared to the hadronic sum rule from the bag
  model. Solid blue line: estimate from conventional mesons and baryons
  ($C=0$). Dashed green line: estimate when exotic hadrons are included,
  applying a shift $C=-40\MeV$.}
\label{fig:4}
\end{figure}

In Fig.~\ref{fig:4} we display lattice data for $T \log L(T)$ vs the hadronic
sum rule. The hadronic estimate (with no additional finite renormalization in
the case of the bag model) describes well the various lattice data sets in the
range $150\MeV<T<190\MeV$, and even lower temperature data from
\cite{Bazavov:2011nk}. However, the steeper slope displayed by the stout
action data \cite{Borsanyi:2010bp} for $T<150\MeV$ cannot possibly be
saturated with conventional mesons and hadrons, since all these states have
already been accounted for. Inclusion of exotic hadrons, $hq\bar{q}\bar{q}$
(tetraquarks) and $hqqq\bar{q}$ (pentaquarks), do actually produce the
required slope (see Fig.~\ref{fig:4}), and it would imply a much shorter
temperature range of applicability of the hadronic sum rule. At present, the
various sets of lattice data disagree at the lowest temperatures, and also it
is unsettled whether exotic hadrons are present in the QCD spectrum or not
\cite{Jaffe:2004ph}. Our analysis implies that resolution of one of these
issues would shed light on the other.

{\em Final comments.---}From QCD considerations, we derive a Boltzmann
distribution formula in terms of hadrons for the expectation value of the
Polyakov loop in the confined phase, as required by quark-hadron duality,
where its real and positive character becomes manifest. Our derivation exposes
the obvious fact that the Polyakov loop gets its expectation value from the
dressing with dynamical quarks or antiquarks. Since $\U(N_f)$ is an exact
global symmetry, the numerator in \Eq{31}, and hence $L(T)$ itself, can be
decomposed into separate contributions from different flavors and different
baryon numbers. Such decompositions are in principle accessible to lattice
calculations (although with a difficulty similar to that of introducing a
chemical potential) and they would provide further information about the
interplay between the QCD thermal state and the heavy quark spectrum.

\begin{acknowledgments}We thank C. Garcia-Recio for comments on the manuscript.
This work has been supported by Plan Nacional de Altas Energ\'{\i}as
(FPA2008-01430 and FPA2011-25948), DGI (FIS2011-24149), Junta de
Andaluc{\'\i}a grant FQM-225, and the Spanish Consolider-Ingenio 2010
Programme CPAN (CSD2007-00042). E.M. would like to thank the Institute
for Nuclear Theory at the University of Washington, USA, and the Institut
f\"ur Theoretische Physik at the Technische Universit\"at Wien, Austria, for
their hospitality and partial support during the completion of this work. The
research of E.~Meg\'{\i}as is supported by the Juan de la Cierva Program of
the Spanish MICINN.
\end{acknowledgments}

%\bibliography{refs}


\end{document}